\documentclass[conference]{IEEEtran}
\IEEEoverridecommandlockouts

\usepackage{cite}
\usepackage{amsmath,amssymb,amsfonts}
\usepackage{url}
\usepackage{algorithmic}
\usepackage{graphicx}
\usepackage{textcomp}
\usepackage{xcolor}
\usepackage{colortbl}
\usepackage{color}
\usepackage{booktabs,tabulary}
\usepackage{multirow}
\usepackage{adjustbox}
\usepackage{fancyhdr}
\usepackage{rotating}
\usepackage{amsmath}
\usepackage{color,soul}
\usepackage{siunitx}
\usepackage{amssymb}
\usepackage{booktabs,tabulary}
\usepackage{tabularx}
\usepackage{float}
\usepackage{notoccite}
\usepackage{multirow}
\usepackage{rotating}
\usepackage{tikz}
\usepackage{balance}
\usepackage{listings}
\usepackage{pdfpages}
\usepackage{subcaption}
\usepackage{url}
\usepackage{array}
\usepackage{wrapfig}
\usepackage{tabularx}
\usepackage{booktabs}
\usepackage{multirow}
\usepackage{xcolor}
\usepackage{microtype}

\usepackage{mdframed}
\newmdenv[
  backgroundcolor=gray!10,
  linecolor=gray!50,
  roundcorner=5pt,
  linewidth=0.5pt,
  innertopmargin=6pt,
  innerbottommargin=6pt,
  innerleftmargin=6pt,
  innerrightmargin=6pt
]{roundedbox}

\def\BibTeX{{\rm B\kern-.05em{\sc i\kern-.025em b}\kern-.08em
    T\kern-.1667em\lower.7ex\hbox{E}\kern-.125emX}}

\begin{document}

\title{STMutants: A Mutation Testing Dataset for Structured Text
Programs in Industrial Automation}


\author{
    \IEEEauthorblockN{Md Humaun Kabir}
    \IEEEauthorblockA{
        Center for Data Analytics and Cybersecurity \\
        Lamar University \\
        Beaumont, TX, USA \\
        Email: mkabir13@lamar.edu
    } \and 
      \IEEEauthorblockN{Md Rakibul Islam}
    \IEEEauthorblockA{
        Department of Computer Science \\
        Lamar University \\
        Beaumont, TX, USA \\
        Email: mislam108@lamar.edu
    }
    \and
    \IEEEauthorblockN{Helen H. Lou}
    \IEEEauthorblockA{
        Dept. of Chem. \& Biomol. Engineering \\
        Lamar University \\
        Beaumont, TX, USA \\
        Email: hhlou@lamar.edu
    }
}

\maketitle

\begin{abstract}
Mutation testing is widely used to evaluate test-suite effectiveness, yet
IEC~61131-3 Structured Text (ST) programs still lack a publicly
available benchmark that supports reproducible mutation-based research.
This gap is especially important because ST is extensively used in
Programmable Logic Controllers (PLCs) that operate in real-time,
safety-critical industrial environments, where software faults may cause
equipment damage, production loss, or unsafe system behavior. To address
this need, we present \textbf{STMutants}, a curated mutation testing
dataset for industrial automation software. STMutants contains 110
generated first-order mutants derived from 11 ST programs collected from
the OSCAT basic library and industrially relevant sources, of which 108
are retained after observability and equivalence screening. The dataset
covers seven mutation operator categories adapted from classical
taxonomies for the PLC domain, including value, relational, arithmetic,
logical, negation, operation insertion/omission, and initialization
faults. Each mutant is constructed through a four-phase methodology:
fault-type profiling and operator selection, syntactic transformation,
compilability verification, and manual equivalence screening with strong
inter-rater agreement ($\kappa = 0.87$). To demonstrate the usefulness
of the dataset, we evaluate three large language models (LLMs) in a
two-phase setting: test-suite generation followed by mutation
kill/survive prediction. Across 108 retained mutants, the models achieve
mutation detection accuracies of 86.1\%, 94.4\%, and 86.1\%,
respectively, with statistical analysis confirming significant
performance differences. By providing the first publicly available
mutation benchmark for ST programs, STMutants enables reproducible
research on automated test generation, mutation analysis, fault
localization, and AI-assisted quality assurance for PLC software.
\end{abstract}


\section{Introduction}

Programmable Logic Controllers (PLCs) govern critical processes in manufacturing, energy, transportation, and infrastructure systems. Modern PLC applications are increasingly implemented using the IEC~61131-3 Structured Text (ST) language, which supports complex arithmetic operations, state machines, and modular function block compositions executed under strict real-time constraints. Because these systems frequently operate in safety-critical environments, software faults can lead to severe consequences including equipment damage, production loss, or even physical harm~\cite{Leveson1993}. Ensuring the reliability of PLC software is therefore a central challenge in industrial automation.

One widely used technique for evaluating test suite effectiveness is \emph{mutation testing}. Originally proposed by DeMillo et al.~\cite{DMillo1978},~\cite{IslamDBSecQA2026} and Hamlet~\cite{Hamlet1977}, mutation testing introduces small syntactic changes (mutants) into a program and measures whether the existing test suite can detect the injected faults. The method is grounded in two foundational hypotheses: the \emph{Competent Programmer Hypothesis}, which assumes that real faults are typically small deviations from correct programs~\cite{Budd1980}, and the \emph{Coupling Effect}, which states that test cases capable of detecting simple faults are also likely to detect more complex ones~\cite{DMillo1978,O_ffutt1992}. Together these principles justify the use of \emph{first-order mutants}, a design choice widely validated in studies of mainstream programming languages~\cite{Andrews2005,just2014}. Over the past decades, mutation testing has proven effective for evaluating test adequacy in languages such as C, Java, and Python~\cite{just2014,coles2016} ~\cite{11410319}.

Despite these advances, mutation testing research for PLC software—particularly IEC~61131-3 Structured Text—remains largely unexplored. Unlike mainstream software engineering domains, there is currently no publicly available benchmark dataset of ST mutants that researchers and practitioners can use for evaluation and comparison. The absence of such a benchmark limits reproducibility, hinders tool development, and prevents systematic assessment of emerging techniques such as automated testing and large language model (LLM)-based analysis for PLC programs.

To address this gap, we introduce \textbf{STMutants}~\cite{STMutantDataset}, a dataset containing 110 first-order mutants across 11 Structured Text programs drawn from the OSCAT basic library and industrial sources. The dataset is constructed using seven mutation operator categories adapted from classical mutation testing taxonomies~\cite{OFfutt1996,agrawal1989,IslamDBSecQA2026}. Each mutant is generated through a structured four-phase methodology that includes mutation operator selection, syntactic transformation, compilability verification, and equivalence screening. As an initial baseline evaluation, we apply three large language models to a two-phase mutation analysis task: first generating test suites for the programs, and then predicting whether individual mutants are killed or survive. The models achieve mutation prediction accuracies of 86.1\%, 94.4\%, and 86.1\%, respectively, establishing the first reproducible baseline results for mutation testing research in IEC~61131-3 ST programs. The dataset is available in the following link: \textbf{\url{https://doi.org/10.6084/m9.figshare.31724761}}.

\section{Dataset Construction Methodology}

\subsection{Program Selection}

The dataset comprises 11 ST programs listed in
Table~\ref{tab:programs}. Nine programs are drawn from the OSCAT basic
library; two---\textsc{Traffic\_Ctrl} and \textsc{Dec\_To\_Hex}---are
sourced from Koziolek et al.~\cite{HK2024}. Programs span 38--211
lines of code (LOC) and cover five functional categories, ensuring
coverage of key PLC testing challenges: complex control flow, internal
state retention, timer-dependent behavior, and numeric precision.

\begin{table}[h]
\caption{ST Programs in the STMutants Dataset}
\vspace{-0.2cm}
\label{tab:programs}
\scriptsize
\setlength{\tabcolsep}{3pt}
\begin{tabularx}{\linewidth}{%
  >{\raggedright\arraybackslash}p{2cm}
  >{\raggedright\arraybackslash}X
  >{\centering\arraybackslash}p{1.2cm}
  >{\centering\arraybackslash}p{0.8cm}
  >{\centering\arraybackslash}p{1cm}}
\toprule
\textbf{Name} & \textbf{Description} & \textbf{Category} & \textbf{LOC} & \textbf{Source} \\
\midrule
CRC\_GEN      & CRC checksum generator          & Logic    & 66  & OSCAT    \\
DEC\_TO\_HEX  & Decimal-to-hex string           & String   & 43  & Koziolek \\
FLOW\_METER   & Flow rate per unit time         & Measure  & 68  & OSCAT    \\
FT\_PIDWL     & PID with wind-up reset          & Control  & 50  & OSCAT    \\
GEN\_BIT      & Binary signal pattern generator & Pulse    & 80  & OSCAT    \\
GEN\_SIN      & Sine wave generator             & Signal   & 46  & OSCAT    \\
LAMBERT\_W    & Lambert W function              & Math     & 38  & OSCAT    \\
MATRIX        & 4$\times$5 keyboard matrix      & Logic    & 103 & OSCAT    \\
SEQUENCE\_8   & 8-bit signal sequencer          & Pulse    & 211 & OSCAT    \\
TOOL\_CHANGER & Tool carousel controller        & Discrete & 45  & OSCAT    \\
TRAFFIC\_CTRL & Traffic light controller        & Discrete & 94  & Koziolek \\
\bottomrule
\end{tabularx}
\vspace{-0.4cm}
\end{table}

\subsection{Mutation Operator Categories}
\label{sec:operators}

We define seven mutation operator categories for IEC~61131-3 ST,
adapted from classical taxonomies~\cite{agrawal1989,Ma2005,harman2011}.
The seven categories collectively cover all major syntactic fault
classes in ST programs---value faults (VRO), condition faults
(CRO, LRO, NIO), computation faults (ARO, AOIO), and state
initialization faults (IO)---satisfying the sufficient mutation
operator principle of Offutt et al.~\cite{OFfutt1996} while
remaining tractable for manual equivalence screening. Each category
targets fault types empirically observed in industrial control
software.

\begin{sloppypar}

\textbf{Value Replacement Operator (VRO):}
Replaces a numeric constant, boolean literal, or time literal with a
semantically close alternative, simulating off-by-one errors, incorrect
threshold values, and initialization mistakes prevalent in real-time
controllers. Numeric constants are perturbed by the smallest
semantically meaningful delta; boolean constants are toggled
(\texttt{TRUE}~$\leftrightarrow$~\texttt{FALSE}); time literals are
changed to the nearest distinct time unit (e.g., \texttt{t\#0s} to
\texttt{t\#0ms}). Corresponds to the Constant Replacement operator of
Agrawal et al.~\cite{agrawal1989} and the Scalar Variable Replacement
operator of Offutt et al.~\cite{OFfutt1996}.

\textbf{Comparison/Relational Operator (CRO):}
Replaces a relational operator with an adjacent one that alters
boundary semantics (e.g., $<$ to $\leq$; $=$ to $\geq$), targeting
boundary conditions critical in time-dependent controllers. The
replacement is always the closest operator in semantic distance to
minimize trivial detection. Corresponds to the Relational Operator
Replacement (ROR) operator~\cite{agrawal1989,OFfutt1996}.

\textbf{Arithmetic Replacement Operator (ARO):}
Replaces an arithmetic operator with a semantically distinct
counterpart (e.g., $\times$~to~$+$), targeting computational
correctness in signal processing and numeric control logic. Models
common computation faults including sign errors, incorrect accumulation,
and wrong scaling relationships~\cite{agrawal1989}.

\textbf{Negation Insertion/Omission (NIO):}
Inserts a \texttt{NOT} operator before a boolean sub-expression to
invert its logic, or removes an existing \texttt{NOT} to restore
uninverted behavior. Parentheses are adjusted where required for
syntactic validity. Inverting guard conditions can silence safety
interlocks---a particularly dangerous fault class in alarm logic~\cite{agrawal1989}.

\textbf{Logical/Boolean Operator (LRO):}
Replaces a boolean connective with its dual
(e.g., \texttt{AND}~$\leftrightarrow$~\texttt{OR}) in compound
conditions, potentially reversing control-flow semantics entirely.
Corresponds to the Logical Operator Replacement (LOR)
operator~\cite{agrawal1989}. Such faults frequently propagate to
system failures in safety-critical applications~\cite{Lutz1993}.

\textbf{Arithmetic Operation Insertion/Omission (AOIO):}
Inserts an arithmetic operation into an expression (e.g., \texttt{pt}
changed to \texttt{pt/2}), or omits an existing one (e.g., removing a
scaling step from a composite expression). Introduces incorrect signal
processing, timing errors, and control
instabilities~\cite{agrawal1989}.

\textbf{Initialization Operator (IO):}
Changes a variable initialization assignment in the declaration block
to use the opposite default value (e.g.,
\texttt{significantDigitFound~:=~FALSE} to \texttt{:=~TRUE}).
Represents incorrect default state assumptions, corresponding to
Variable Initialization Fault patterns in fault injection
research~\cite{Arlat1990}. Initialization errors cause systems to
start in incorrect states with unpredictable behavior in cyclic PLC
execution environments~\cite{Frey2000}.

\end{sloppypar}

\subsection{Mutation Generation Methodology}
\label{sec:mutation-methodology}

Generating a reproducible mutation dataset requires careful control
over the mutation process to ensure that (i)~mutants faithfully
represent plausible faults, (ii)~all mutants are syntactically valid
and compilable, and (iii)~equivalent mutants are excluded. We follow a
structured four-phase process. Note that Phase~1 references the operator
definitions in Section~\ref{sec:operators} above.

\subsubsection{\textbf{Phase 1:} Fault-Type Profiling and Operator Selection}

Before generating any mutant, each ST program was analyzed to identify
its \emph{mutation-susceptible locations}: points in the source code
where a single syntactic change could plausibly represent a realistic
programmer fault.

\textbf{Structural Parsing.} Each ST program is parsed into its
constituent statement types: assignment statements, conditional
branches (\texttt{IF}/\texttt{ELSIF}/\texttt{CASE}), loop constructs
(\texttt{FOR}/\texttt{WHILE}/\texttt{REPEAT}), function block
instantiation, and timer/counter function calls. For each statement,
we identify all operator tokens, operand constants, and boolean
sub-expressions as candidate mutation sites.

\textbf{Operator Applicability Mapping.} Each operator category
(Section~\ref{sec:operators}) is mapped to the statement types where
it is semantically applicable. CRO applies exclusively to relational
sub-expressions ($<$, $\leq$, $>$, $\geq$, $=$, $\neq$); VRO applies
to numeric and boolean literal constants; LRO applies to boolean binary
operators (\texttt{AND}, \texttt{OR}, \texttt{XOR}); NIO targets
\texttt{NOT} prefixes on boolean expressions; ARO applies to arithmetic
binary operators ($+$, $-$, $\times$, $\div$, \texttt{MOD}); AOIO
targets entire arithmetic sub-expressions; and IO addresses
initialization assignments in variable declaration blocks.

\textbf{Mutation Point Enumeration.} All applicable mutation points
are enumerated and recorded, forming the \emph{mutation candidate set}
$\mathcal{C}(P)$ for program $P$.

\textbf{Stratified Sampling.} To ensure operator diversity within the
fixed budget of 10 mutants per program, operator categories are
weighted proportional to their representation in $\mathcal{C}(P)$,
subject to a minimum of one mutant per applicable operator category
when the budget permits. The stratified sample is then subject to
manual review to confirm that each selected mutation point represents
a plausible real-world fault.

\subsubsection{\textbf{Phase 2:} Syntactic Transformation}

Each selected mutation point is transformed by applying a single
atomic token substitution per the operator rules in
Section~\ref{sec:operators}. Only the identified token at the mutation
point is replaced; all surrounding code is left byte-for-byte
unchanged. Each transformed source file constitutes one mutant file,
named \texttt{1.txt} through \texttt{10.txt} within each program
folder, alongside a mutation description document recording the
operator type, source line, original token, replacement token, and
fault rationale.

\subsubsection{\textbf{Phase 3:} Compilability Verification}

Every generated mutant is verified for syntactic correctness by
compiling with \texttt{MATIEC}~\cite{MATIEC}, a widely used IEC~61131-3-compliant
development environment. Compilation verifies syntactic and
type-level validity; it does not guarantee behavioral divergence from
the original, which is addressed by the equivalence screening in
Phase~4. Mutants that fail compilation are discarded and replaced by
the next candidate from $\mathcal{C}(P)$. In practice, fewer than 5\%
of generated candidates were rejected at this stage, as the operator
applicability mapping in Phase~1 ensures type safety by construction
for most transformations.

\subsubsection{\textbf{Phase 4:} Equivalence Screening}

Equivalent mutants degrade mutation score validity by inflating the
denominator without representing genuine faults. We apply a two-step
screening process.

\textbf{Step 1: Static Pre-Screening}. Before inter-rater review, each mutant was
examined by one author through direct reading of the mutated source
code to identify two categories of likely equivalence. First,
\emph{constant folding candidates}: mutants where the changed token
appears inside a branch or expression whose value can be determined
at inspection time to be unreachable or invariant under all plausible
inputs (e.g., a constant modified inside a dead-code branch). Second,
\emph{redundant assignment candidates}: mutants where the modified
variable is immediately overwritten before any read, rendering the
mutation unobservable regardless of input. Both categories were
identified by manually tracing the data-flow of the mutated variable
from its mutation site forward through the program text to its next
read or output. Mutants flagged under either category were passed to
the full two-author independent review described below.

\textbf{Step 2: Independent Dual-Review} Two reviewers with over two years of ST programming experience independently reviewed each
flagged candidate, as well as a random 20\% sample of all non-flagged
mutants, recording a binary judgment (equivalent / non-equivalent).
Inter-rater agreement was measured using Cohen's kappa ($\kappa =
0.87$, indicating strong agreement~\cite{landis1977}). Any candidate
judged equivalent by at least one reviewer was excluded and replaced
with the next viable candidate from $\mathcal{C}(P)$.

For \textsc{Flow\_Meter} specifically, two of the 10 generated
mutants are found to be \emph{test-unobservable}: their effects do not propagate to any observable output variable under any reachable input, as determined by static data-flow analysis. These two are excluded. Consequently, \textsc{Flow\_Meter}
contributes 8 test-observable mutants rather than 10 to the retained
dataset, and its row in Table~\ref{tab:results} reports results over
$N{=}8$ rather than $N{=}10$. The operator distribution row in
Table~\ref{tab:operators} reflects the 10 \emph{generated} mutants
prior to observability filtering; the two excluded mutants were an
ARO and an AOIO instance. The total retained dataset comprises
\textbf{108 test-observable mutants} across 11 programs.
\subsubsection{Rationale for 10 Mutants per Program}

The choice of exactly 10 mutants per program is motivated by three
considerations. First, it provides a uniform, directly comparable
baseline across all 11 programs regardless of LOC. Second, it aligns
with the ``sufficient mutant'' principle of Offutt et
al.~\cite{OFfutt1996}, which shows that a small, diverse set of
operators achieves near-complete fault detection relative to the full
mutation set. Third, the 10-mutant budget is sufficient to instantiate
all applicable operator categories for each program, ensuring operator
diversity without oversampling any single fault type.

\subsection{Mutation Strategy}
\label{sec:mutation-strategy}

Following systematic mutation testing practice~\cite{OFfutt1996}, we
generate exactly 10 mutants per program (110 generated; 108
test-observable). All mutants are \emph{first-order}
mutations---single syntactic changes---consistent with established
findings that first-order mutations are sufficient for effective test
evaluation~\cite{harman2011,O_ffutt1992,kamal2025robust}. All retained mutants are
syntactically valid, compilable, and non-equivalent following the
screening process in Section~\ref{sec:mutation-methodology}.

\subsection{Theoretical Justification for First-Order Mutations}

\textbf{Sufficient Mutation Operators:} Offutt et al.~\cite{OFfutt1996}
demonstrated that a reduced set of carefully selected mutation operators
can achieve nearly the same fault detection effectiveness as the full
mutation set while significantly reducing computational cost. Our seven
operator categories align with this principle.

\textbf{Empirical Validation:} Empirical studies by Offutt et al.\ and
Jia and Harman~\cite{O_ffutt1992,harman2011} have shown that
first-order mutations are sufficient for effective testing, with
higher-order mutations providing marginal additional benefit at
substantially increased cost. This supports our exclusive focus on
first-order mutations.

\section{Dataset Description}
\label{sec:dataset-description}

\subsection{Operator Distribution and Statistics}

Table~\ref{tab:operators} shows the full distribution of the seven
mutation operators across all 11 programs. In total, 110 mutants were
generated; 108 are retained after excluding 2 test-unobservable
mutants from \textsc{Flow\_Meter} (see Section~\ref{sec:mutation-methodology},
Phase~4).

\textbf{VRO} is the most prevalent category at 41 generated mutants
(37.3\%), followed by \textbf{CRO} at 36 (32.7\%), \textbf{ARO} at 12 generated mutants
(10.9\%), \textbf{NIO} at 10 (9.1\%), \textbf{LRO} at 5 (4.5\%),
\textbf{AOIO} at 4 (3.6\%), and \textbf{IO} at 2 (1.8\%). VRO and CRO
together account for exactly 70.0\% of all generated mutants (77 of
110). This distribution reflects the syntactic composition of the
selected programs: the OSCAT programs are numerically intensive
function blocks with many literal constants and relational guards, which
naturally yield a large candidate set $\mathcal{C}(P)$ for VRO and CRO
operators under the stratified sampling scheme. Whether this distribution
generalizes to industrial PLC codebases at scale remains an open
empirical question and is acknowledged in Section~\ref{sec:threats}.
The IO operator is underrepresented (2 mutants) because most OSCAT
programs do not expose mutation-susceptible initialization sites; this
reflects the actual fault space of the selected programs rather than an
intentional design choice, and per-operator conclusions for IO should
be treated as illustrative.


\begin{figure*}[htb]
  \centering
  \includegraphics[width=0.55\textwidth, height=0.50\textheight]{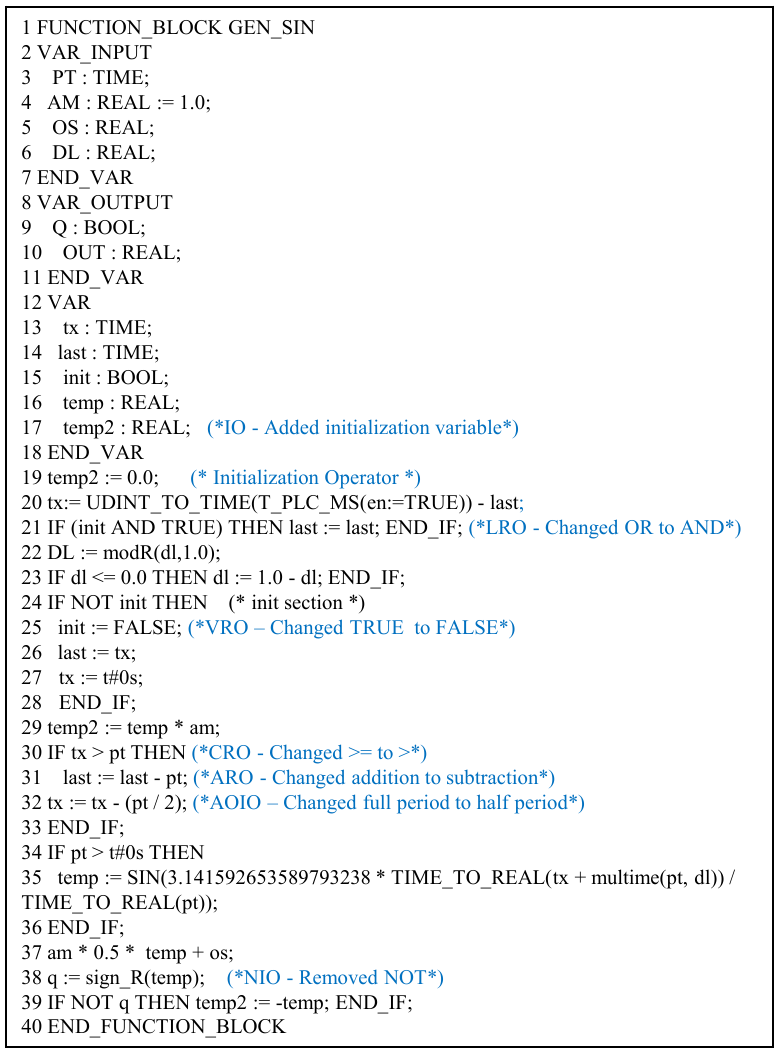}
    \caption{Annotated first-order mutants applied to the
  \textsc{Gen\_Sin} ST function block. Operators shown: IO (line~19),
  LRO (line~21), VRO (line~25), CRO (line~30), ARO (line~31),
  AOIO (line~32), and NIO (line~38).}
  \vspace{-0.5cm}
  \label{fig:gen_sin_mutations}
    \vspace{-0.1cm}
\end{figure*}



\begin{table}[h]
\captionsetup{font=footnotesize}
\centering
\caption{Distribution of Generated Mutation Operators Across Programs.
  $^\dagger$FLOW\_METER: 10 mutants generated; 2 excluded as
  test-unobservable (1~ARO, 1~AOIO), yielding 8 retained.}
\label{tab:operators}
\vspace{-0.2cm}
\fontsize{6.8}{8.2}\selectfont
\setlength{\tabcolsep}{2.2pt}
\begin{tabular}{
>{\raggedright\arraybackslash}p{1.5cm}
>{\centering\arraybackslash}p{0.42cm}
>{\centering\arraybackslash}p{0.48cm}
>{\centering\arraybackslash}p{0.42cm}
>{\centering\arraybackslash}p{0.42cm}
>{\centering\arraybackslash}p{0.42cm}
>{\centering\arraybackslash}p{0.3cm}
>{\centering\arraybackslash}p{0.42cm}
>{\centering\arraybackslash}p{0.45cm}
}
\toprule
\textbf{Program} & \textbf{ARO} & \textbf{AOIO} & \textbf{CRO} & \textbf{VRO} & \textbf{LRO} & \textbf{IO} & \textbf{NIO} & \textbf{Gen.} \\
\midrule
CRC\_GEN        & 2 & 1 & 4 & 2 &   & 1 &   & 10 \\
DEC\_TO\_HEX    & 3 & 2 & 4 & 1 &   &   &   & 10 \\
FLOW\_METER$^\dagger$ & 1 & 1 & 1 & 4 & 1 & 1 & 1 & 10 \\
FT\_PIDWL       &   &   & 4 & 3 &   &   & 3 & 10 \\
GEN\_BIT        &   &   & 6 & 2 & 2 &   &   & 10 \\
GEN\_SIN        & 4 &   & 2 & 4 &   &   &   & 10 \\
LAMBERT\_W      & 2 &   & 2 & 6 &   &   &   & 10 \\
MATRIX          &   &   & 1 & 8 & 1 &   &   & 10 \\
SEQUENCE\_8     &   &   & 7 & 2 &   &   & 1 & 10 \\
TOOL\_CHANGER   &   &   & 4 & 2 &   &   & 4 & 10 \\
TRAFFIC\_CTRL   &   &   & 1 & 7 & 1 &   & 1 & 10 \\
\midrule
\textbf{Total (gen.)} & \textbf{12} & \textbf{4} & \textbf{36} & \textbf{41} & \textbf{5} & \textbf{2} & \textbf{10} & \textbf{110} \\
\textbf{Retained}     & \textbf{11} & \textbf{3} & \textbf{36} & \textbf{41} & \textbf{5} & \textbf{2} & \textbf{10} & \textbf{108} \\
\bottomrule
\end{tabular}
\vspace{-0.3cm}
\end{table}

\subsection{Representative Mutation Examples}

Figure~\ref{fig:gen_sin_mutations} illustrates seven first-order mutants
applied to the \textsc{Gen\_Sin} function block. Each red annotation
marks a single injected token at its mutation site: an IO mutation
introduces a spurious initialization assignment at line~19
(\texttt{temp2~:=~0.0}); an LRO mutation replaces \texttt{OR} with
\texttt{AND} in the guard condition at line~21; a VRO mutation toggles
\texttt{init} from \texttt{TRUE} to \texttt{FALSE} at line~25; a CRO
mutation changes \texttt{>=} to \texttt{>} at the cycle-boundary check
at line~30; an ARO mutation changes addition to subtraction at
line~31, corrupting the period accumulator; an AOIO mutation halves
the period by replacing \texttt{pt} with \texttt{pt/2} at line~32;
and a NIO mutation removes the \texttt{NOT} operator from the boolean
output assignment at line~38. Together these examples demonstrate how
a single token substitution at a critical site can silently alter
timing behavior, invert guard logic, or corrupt numerical
output---the fault classes most prevalent in the STMutants dataset.



\subsection{Formal Definitions}

Let $P$ be an ST program, $T$ a test suite, and $M$ the set of all
test-observable, non-equivalent mutants generated from $P$ via the
four-phase process of Section~\ref{sec:mutation-methodology}. A mutant
$m(P) \in M$ is produced by applying a single mutation operator $\mu
\in \Omega$ (where $\Omega$ is our set of seven operator categories) to
a single mutation point in $P$. A mutant $m$ is \emph{killed} if there
exists at least one test case $t \in T$ such that $t(P) \neq t(m(P))$.
A mutant is \emph{equivalent} if $\forall\, t \in T,\; t(P) = t(m(P))$;
equivalent mutants are excluded from $M$.

The \emph{mutation score} $\mathit{MS}(T, M)$ is:
\[
  \mathit{MS}(T, M) \;=\;
  \frac{\bigl|\{ m \in M \mid \exists\, t \in T : t(P) \neq t(m(P))
  \}\bigr|}{|M|}
\]
A higher mutation score indicates greater test suite effectiveness at
detecting the fault types modelled by the operator set $\Omega$.

\section{Experiment with the Dataset}
In the following, we evaluate the STMutants dataset through a
controlled experiment with three LLMs, guided by the following
research question (RQ).

\textit{RQ: To what extent do LLM-generated unit test suites detect
mutations in the STMutants dataset, and how does detection performance
vary across mutation operator categories?}

This RQ assesses LLM-generated test suites' ability to kill mutants,
accounting for operator-specific challenges (e.g., boundary conditions
in CRO, boolean inversions in NIO) and program attributes such as
cyclomatic complexity and temporal dependencies, which may affect
reasoning and coverage in safety-critical control logic.

\subsection{Experimental Setup and LLM Versions}

We evaluate three LLMs as automated mutation analysis tools:
\textbf{GPT-5.2 (gpt-5.2
)}~\cite{openai2025gpt53}, a large-scale autoregressive
model by OpenAI optimized for complex reasoning tasks;
\textbf{Gemini~2.5 (gemini-2.5-pro)}~\cite{google2025gemini31}, an
efficiency-oriented multimodal model by Google DeepMind designed for
fast inference at scale; and \textbf{Claude Sonnet~4.5 (claude-sonnet-4-5-20250929}~\cite{anthropic2025claude46},
a balanced instruction-following model by Anthropic targeting both
reasoning quality and deployment efficiency.  All three models are
queried via their respective APIs using default inference parameters,
e.g., temperature, \texttt{top\_p},
\texttt{top\_k}, and \texttt{max\_tokens}, left at provider defaults.

\subsection{Experiment Design}



\begin{figure}[htbp]
  \centering
      \vspace{-0.3cm}
  \includegraphics[width=1\linewidth, height=0.35\textheight]{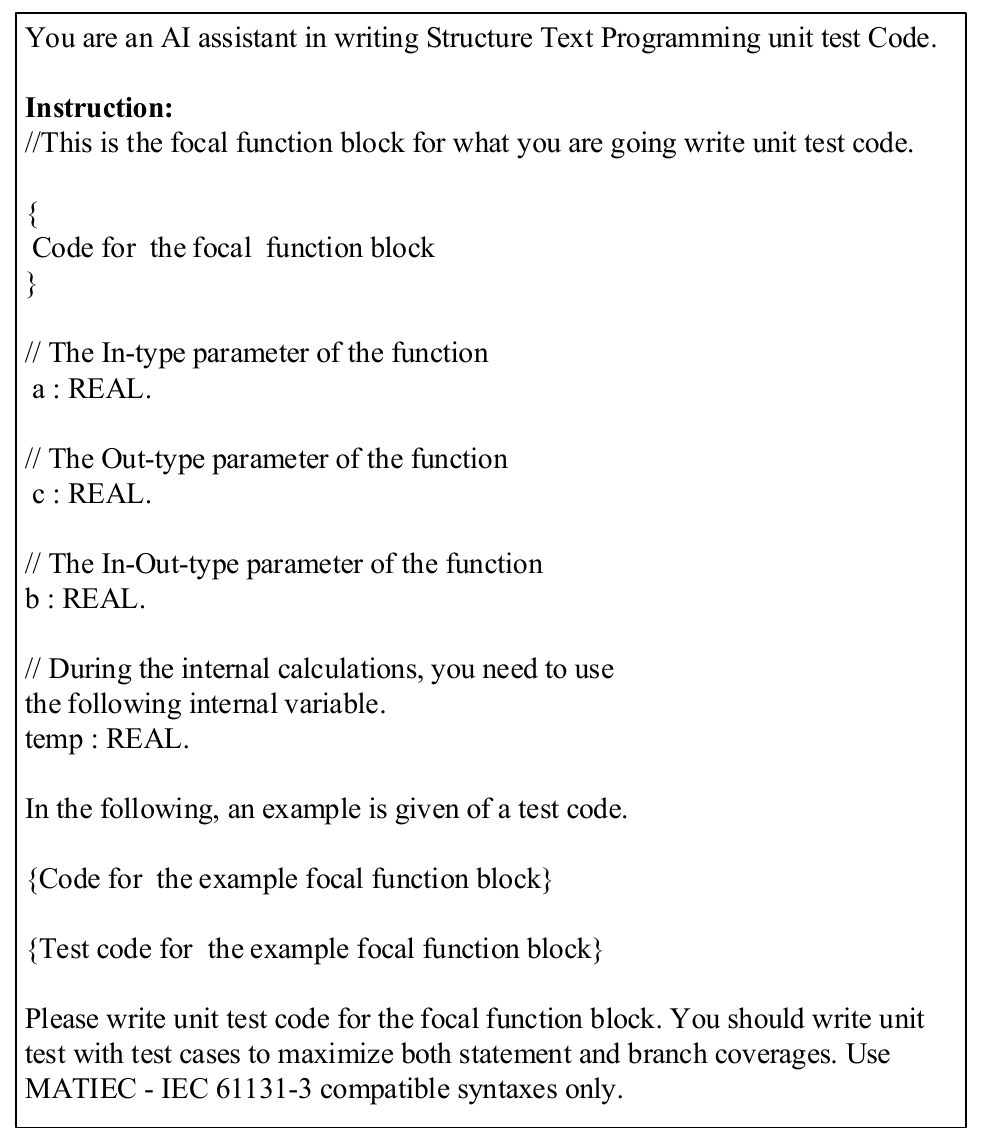}
  \vspace{-0.3cm}
  \captionsetup{font=footnotesize}
    \vspace{-0.3cm}
  \caption{Standardized natural-language prompt used in Phase~A (test
  suite generation). All three models received the same template to
  ensure comparability.}
  \vspace{-0.2cm}
  \label{fig:prompt}
  
\end{figure}

The evaluation proceeds in two distinct phases. In \textbf{Phase~A
(Test Suite Generation)}, each LLM is prompted to generate a unit
test suite for each of the 11 ST programs using a standardized prompt
with a one-shot example, as illustrated in Figure~\ref{fig:prompt}.
The generated test suite specifies input vectors and expected output
values for each function block. One-shot prompting has been shown to
improve LLM performance on code-related tasks~\cite{brown2020gpt3}.

In \textbf{Phase~B (Mutation-Based Evaluation)}, we assess whether the generated test suites are capable of detecting injected faults. Unlike Phase~A, no additional prompt is constructed for this stage. Instead, the test suite generated in Phase~A was directly reused. For each of the 108 retained mutants, the original function block in the test harness was replaced with the corresponding mutated version while keeping the same test inputs and expected outputs.

The test suite was then executed using the \texttt{MATIEC} compiler~\cite{MATIEC} for both the original and mutated programs. A mutant was considered \emph{killed} if executing the same test inputs on the mutant produced outputs that differed from the expected outputs defined in the test suite. Otherwise, the mutant was labeled as \emph{surviving}.

All experiments use one-shot prompting without chain-of-thought
instructions. While this may underestimate the models' full
capabilities, it reflects a realistic deployment scenario. Future work will explore
chain-of-thought and few-shot variants as upper-bound baselines.

\subsection{Evaluation Metrics}

\textbf{Mutation Kill Rate (\%).} The percentage of retained mutants
correctly identified as killed by the LLM-generated test suite. Higher
rates indicate stronger reasoning ability or better test coverage.

\textbf{Mutation Score.} The aggregate percentage of retained mutants
killed across all programs. A surviving mutant represents either a gap
in the test suite or a failure by the LLM to reason about behavioral
differences.

\textbf{Per-Operator Kill Rate.} Disaggregated kill rates by operator
category, directly addressing the operator-specific component of the RQ.

\subsection{Statistical Analysis} We compare mutation detection performance across three LLMs using
{Cochran's Q test}~\cite{cochran1950}---the appropriate non-parametric test for
comparing three or more matched binary outcomes on the same set of
instances. A significant Cochran's Q result is
followed by pairwise {Wilcoxon signed-rank}~\cite{wilcoxon1945} post-hoc tests
with {Holm--Bonferroni}~\cite{holm1979} correction to control the family-wise
error rate across three pairwise comparisons.

\subsection{Results}
\label{sec:results}

As shown in Table~\ref{tab:results}, over 108 retained mutants, \textbf{Gemini~2.5} achieves the
highest mutation score at \textbf{94.4\%} (102 killed, 6 missed),
followed by \textbf{GPT-5.2} and \textbf{Claude Sonnet~4.5}, both at
\textbf{86.1\%} (93 killed, 15 missed each). Cochran's Q test yields
$Q = 8.17$ ($p = 0.017$), confirming that the three models do not
perform equivalently. Pairwise Wilcoxon signed-rank post-hoc tests
with Holm--Bonferroni correction show that Gemini~2.5
significantly outperforms both GPT-5.2 ($p = 0.029$) and Claude
Sonnet~4.5 ($p = 0.029$), while GPT-5.2 and Claude Sonnet~4.5 are
not significantly different ($p = 1.00$ after correction).

\begin{table}[h]
\centering
\captionsetup{font=footnotesize}
\caption{Mutation Kill/Survive Results per LLM (over 108 retained
  mutants; FLOW\_METER $N{=}8$, all others $N{=}10$).
  Claude~S.\ = Claude Sonnet~4.5.}
\vspace{-0.2cm}
\label{tab:results}
\fontsize{7}{8.5}\selectfont
\setlength{\tabcolsep}{2pt}
\begin{tabular}{%
  >{\raggedright\arraybackslash}p{2cm}
  >{\raggedright\arraybackslash}p{0.45cm}
  >{\raggedright\arraybackslash}p{0.45cm}
  >{\raggedright\arraybackslash}p{0.45cm}
  >{\raggedright\arraybackslash}p{0.45cm}
  >{\raggedright\arraybackslash}p{0.45cm}
  >{\raggedright\arraybackslash}p{0.45cm}
  >{\raggedright\arraybackslash}p{0.35cm}}
\toprule
 & \multicolumn{2}{l}{\textbf{GPT-5.2}} &
   \multicolumn{2}{l}{\textbf{Gemini~2.5}} &
   \multicolumn{2}{l}{\textbf{Claude~S.}} & \\
\cmidrule(lr){2-3}\cmidrule(lr){4-5}\cmidrule(lr){6-7}
\textbf{Program} & Kill & Miss & Kill & Miss & Kill & Miss & \textbf{N} \\
\midrule
CRC\_GEN      &  9 & 1 & 10 & 0 & 10 & 0 & 10 \\
DEC\_TO\_HEX  &  9 & 1 &  9 & 1 &  9 & 1 & 10 \\
FLOW\_METER   &  5 & 3 &  7 & 1 &  8 & 0 &  8 \\
FT\_PIDWL     &  9 & 1 &  9 & 1 &  9 & 1 & 10 \\
GEN\_BIT      &  9 & 1 &  9 & 1 & 10 & 0 & 10 \\
GEN\_SIN      &  8 & 2 &  8 & 2 &  8 & 2 & 10 \\
LAMBERT\_W    & 10 & 0 & 10 & 0 &  8 & 2 & 10 \\
MATRIX        & 10 & 0 & 10 & 0 & 10 & 0 & 10 \\
SEQUENCE\_8   &  5 & 5 &  1 & 9 &  1 & 9 & 10 \\
TOOL\_CHANGER &  9 & 1 &  9 & 1 & 10 & 0 & 10 \\
TRAFFIC\_CTRL & 10 & 0 & 10 & 0 & 10 & 0 & 10 \\
\midrule
\textbf{Total} & \textbf{93} & \textbf{15} & \textbf{102} & \textbf{6}
  & \textbf{93} & \textbf{15} & \textbf{108} \\
\textbf{Score} & \multicolumn{2}{l}{\textbf{86.1\%}} &
                 \multicolumn{2}{l}{\textbf{94.4\%}} &
                 \multicolumn{2}{l}{\textbf{86.1\%}} & \\
\bottomrule
\end{tabular}
\vspace{-0.3cm}
\end{table}

\textbf{Per-operator analysis.} Table~\ref{tab:operator_kill} breaks
down kill rates by operator category. All three models achieve perfect
kill rates on ARO, AOIO, and IO operators. CRO is well-handled by
Gemini~2.5 (100.0\%) but notably lower for Claude Sonnet~4.5
(80.6\%), driven by surviving mutants in \textsc{Sequence\_8}.
NIO proves moderately challenging across all models
(70.0\%--90.0\%), while LRO shows divergent behavior: Claude
Sonnet~4.5 kills all five (100.0\%) versus 80.0\% for the other
two. VRO mutations are the hardest category overall (82.9\%,
90.2\%, and 85.4\% for GPT-5.2, Gemini~2.5, and Claude Sonnet~4.5
respectively), requiring precise expected-value assertions rather
than simple boundary checks.

\begin{table}[h]
\centering
\captionsetup{font=footnotesize}
\caption{Per-Operator Kill Rates (\%) Across LLMs.
  $n$ = number of retained mutants per operator.
  Claude~S.\ = Claude Sonnet~4.5.}
\label{tab:operator_kill}
\vspace{-0.2cm}
\fontsize{7}{8.5}\selectfont
\setlength{\tabcolsep}{3pt}
\begin{tabular}{
>{\raggedright\arraybackslash}p{0.8cm}
>{\centering\arraybackslash}p{0.3cm}
>{\centering\arraybackslash}p{1.1cm}
>{\centering\arraybackslash}p{1.2cm}
>{\centering\arraybackslash}p{1.1cm}
}
\toprule
\textbf{Op.} & \textbf{$n$} & \textbf{GPT-5.2} & \textbf{Gemini 2.5} & \textbf{Claude~S.} \\
\midrule
VRO   & 41 & 82.9\% & 90.2\% & 85.4\% \\
CRO   & 36 & 88.9\% & 100.0\% & 80.6\% \\
ARO   & 11 & 100.0\% & 100.0\% & 100.0\% \\
NIO   & 10 & 70.0\% & 90.0\% & 80.0\% \\
LRO   &  5 & 80.0\% & 80.0\% & 100.0\% \\
AOIO  &  3 & 100.0\% & 100.0\% & 100.0\% \\
IO    &  2 & 100.0\% & 100.0\% & 100.0\% \\
\midrule
\textbf{All} & \textbf{108} & \textbf{86.1\%} & \textbf{94.4\%} & \textbf{86.1\%} \\
\bottomrule
\end{tabular}
\vspace{-0.5cm}
\end{table}

\textbf{Program-level analysis.} All three models perform perfectly on
\textsc{Matrix} and \textsc{Traffic\_Ctrl}---programs with
well-defined boundary conditions amenable to relational reasoning. In
contrast, \textsc{Sequence\_8} (211~LOC, 8-bit sequencer with deeply
nested temporal state transitions) yields dramatically lower scores:
GPT-5.2 at 50\%, and both Gemini~2.5 and Claude Sonnet~4.5 at
only 10\%. The 9 surviving mutants in \textsc{Sequence\_8} for Gemini
and Claude are concentrated in the CRO operator (7 of 10 generated
mutants) and the single NIO instance, all located in deeply nested
state-transition guards where fault propagation requires simulating
multiple sequential scan cycles before a divergent output is produced.
GPT-5.2's comparatively higher score (50\%) may reflect a more
conservative reasoning strategy under uncertainty, tentatively
predicting ``kill'' for boundary-altering CRO mutations even without
full multi-cycle trace simulation; this hypothesis merits systematic
investigation in future work.

This degradation is not simply a function of program size. It reflects
a fundamental difficulty with \emph{temporal sequential reasoning}:
correctly predicting whether a test case produces divergent output
requires tracking state variable evolution across scan-cycle
iterations and understanding how a mutation at a deeply nested
boundary propagates through the output sequence. This finding indicates
that LLM mutation analysis in its current one-shot form is unreliable
for programs where fault propagation is temporally delayed or
state-dependent---a critical limitation for PLC quality assurance.

\section{Research Opportunities}

STMutants opens several research directions spanning automated testing,
AI-assisted quality assurance, and fault analysis for industrial control
software.

\subsection{Automated Test Generation for PLC Programs}

The dataset enables systematic evaluation of automated test
generation techniques for ST programs. Traditional approaches such as
search-based software testing (SBST)~\cite{harman2011} can use mutation
score as an optimization objective to evolve input sequences that
maximize fault detection. However, unlike conventional software,
PLC programs execute cyclically and maintain persistent state across
scan cycles. This introduces unique challenges in designing search
strategies that must reason about temporal behavior and delayed fault
propagation. STMutants provides a controlled environment for
developing and benchmarking such PLC-aware SBST algorithms.

\subsection{AI-Assisted Test Generation and Program Reasoning}

The baseline evaluation in this paper demonstrates that large language
models (LLMs) can generate test suites capable of detecting a majority
of injected faults, but also reveals clear limitations in programs
requiring temporal reasoning. This creates an opportunity to explore
advanced prompting strategies such as chain-of-thought reasoning,
few-shot examples, or retrieval-augmented generation using
IEC~61131-3 documentation. Fine-tuning models on PLC code corpora or
industrial automation libraries such as OSCAT may further improve
their ability to reason about cyclic execution and control logic.
STMutants provides a reproducible benchmark for comparing these
approaches under a consistent evaluation protocol.



\subsection{Fault Localization for PLC Software}

The dataset also supports research on automated fault localization
techniques for ST programs. Each mutant corresponds to a precisely
known fault location, enabling controlled evaluation of
spectrum-based fault localization (SBFL) methods and statistical
debugging approaches. Because PLC programs often involve complex
control logic and persistent internal state, it remains unclear
whether traditional SBFL metrics perform as effectively in this domain
as they do for mainstream programming languages. STMutants offers a
ground-truth dataset for investigating these questions.

\subsection{Benchmarking AI Models and Hybrid Analysis Systems}

Beyond standalone LLM evaluation, STMutants enables the study of
hybrid analysis systems that combine multiple AI models or integrate
LLM reasoning with traditional static and dynamic analysis tools.
For example, ensemble approaches could combine predictions from
multiple models to improve mutation detection accuracy, while
symbolic execution or runtime monitoring could validate predicted
behavioral differences. Such hybrid pipelines may prove particularly
valuable in safety-critical industrial environments where verification
confidence is essential.


\section{Discussion}

Our evaluation demonstrates that state-of-the-art LLMs can serve as
effective mutation analysis tools for ST programs, achieving 86--94\%
detection accuracy over the retained mutant set. The dominance of VRO
and CRO operators (70\% of generated mutants) reflects the syntactic
composition of the selected OSCAT programs, which are numerically
intensive function blocks with many literal constants and relational
guards. The LLMs' strong performance on these categories confirms that
relational and constant reasoning is within current model capabilities
in a one-shot setting.

However, performance degrades significantly on programs with high
cyclomatic complexity and temporal dependencies. The
\textsc{Sequence\_8} results are the most prominent example. We
attribute this to a mismatch between transformer-based token
prediction and the multi-step, stateful simulation required to assess
fault propagation across PLC scan cycles. The statistically
indistinguishable performance of GPT-5.2 and Claude Sonnet~4.5 (both
86.1\%, $p = 1.00$ after correction) suggests these two models occupy
a similar capability tier for ST mutation analysis in one-shot
settings, while Gemini~2.5's advantage is statistically robust
under the Holm--Bonferroni-corrected Wilcoxon post-hoc analysis.

\section{Threats to Validity and Limitations}
\label{sec:threats}

\subsection{Internal Validity}


The selected mutation operators may not capture all possible fault
types in ST programs. We mitigate this by grounding the taxonomy in
established literature~\cite{agrawal1989,harman2011,OFfutt1996} and
adapting operators to IEC~61131-3 constructs. Manual equivalence
screening used documented inter-rater agreement ($\kappa = 0.87$). The
LLM evaluation used one-shot prompting, which may underestimate model
capabilities. All mutants are first-order; higher-order mutations could
expose additional testing gaps. The IO operator is represented by only
two mutants in the dataset; per-operator conclusions for IO should
therefore be treated as illustrative rather than statistically
conclusive.

\subsection{External Validity}

The 11 programs are of moderate size (38--211 LOC) and may not
represent large-scale industrial PLC systems, though they span diverse
functional categories. Programs drawn from the OSCAT library and
academic sources~\cite{HK2024} may not fully reflect proprietary
industrial codebases. The dataset covers only Structured Text; other
IEC~61131-3 languages (Ladder Logic, Function Block Diagram, Sequential
Function Chart) require separate benchmarks. The VRO/CRO-dominant
operator distribution reflects the syntactic characteristics of the
selected programs; formal correlation of this distribution with
field defect data from industrial PLC deployments remains future work.

\subsection{Construct Validity}

Mutation score as a proxy for test suite quality has been extensively
validated~\cite{just2014,O_ffutt1992}. Compilation in \texttt{MATIEC}~\cite{MATIEC}
verifies syntactic and type-level correctness but does not guarantee
runtime behavioral divergence; behavioral non-equivalence is separately
ensured by Phase~4 equivalence screening. The binary kill/survive
classification does not capture fault severity; future work could
address this via weighted mutation scores.

\section{Related Work}

No publicly available benchmark exists for evaluating test suite
quality via mutation testing of IEC~61131-3 Structured Text programs. For mainstream languages, Defects4J~\cite{just2014}
provides 395+ real Java bugs; the Major mutation testing
framework~\cite{just2011major} supports large-scale Java mutation
studies; PIT~\cite{coles2016} offers practical mutation testing with
13 operators. For embedded and safety-critical systems, Frey and
Litz~\cite{Frey2000} studied PLC faults but did not provide a mutation
benchmark; Lutz~\cite{Lutz1993} analyzed spacecraft software faults,
finding fault patterns consistent with our VRO/CRO operator
distribution. Koziolek et al.~\cite{HK2024} evaluated LLMs for PLC
program understanding but did not address mutation testing. Recent
LLM-based test generation work has achieved competitive mutation scores
on Java and Python, but no such evaluation existed for ST programs prior
to this work. STMutants is the first literature-grounded ST mutation
benchmark across diverse industrial program types, and the first to
provide LLM-based mutation analysis baselines for the IEC~61131-3
domain.

\section{Conclusion}
This paper introduced \textbf{STMutants}, the first publicly available
mutation testing dataset for IEC~61131-3 Structured Text (ST)
programs. The dataset addresses a clear gap in PLC software testing
research by providing a reproducible benchmark tailored to industrial
automation software, where reliability is critical and software faults
can have serious operational and safety consequences. STMutants
contains 110 generated first-order mutants across 11 ST programs from
the OSCAT basic library and industrially relevant sources, with 108
retained after observability and equivalence screening. The dataset is
built through a systematic four-phase methodology encompassing
fault-type profiling, mutation operator selection, syntactic
transformation, compilability verification, and manual equivalence
analysis, yielding a high inter-rater agreement of $\kappa = 0.87$.

Our results show that STMutants is not only a curated benchmark
artifact, but also a useful experimental platform for evaluating
emerging AI-based analysis techniques. In particular, the baseline
study with three large language models demonstrates that LLM-generated
test suites can detect a substantial portion of injected faults,
achieving mutation detection accuracies between 86.1\% and 94.4\%.
At the same time, the results reveal an important limitation:
performance drops sharply on temporally complex programs such as
\textsc{Sequence\_8}, where fault propagation depends on stateful,
multi-cycle reasoning. This finding suggests that while current LLMs
are effective for many value- and condition-based mutations, they are
still less reliable for sequential control logic that requires deeper
temporal reasoning. Beyond the immediate baseline results, STMutants establishes a
foundation for future research on automated test generation, mutation
analysis, equivalent mutant detection, fault localization, and
LLM-assisted verification for PLC software. It also opens the door to
higher-order mutation studies and benchmark extensions across other
IEC~61131-3 languages. By releasing a literature-grounded, documented,
and reproducible benchmark for ST mutation testing, we hope to support
more rigorous and comparable evaluation in industrial software testing
research and to accelerate progress toward dependable AI-assisted
quality assurance for programmable logic controllers.

\section{ACKNOWLEDGMENTS}
This research was conducted as part of the Center for Data Analytics and Cybersecurity and was supported in part by the U.S. Department of Energy (DOE) under Grant No. DE-CR0000035. Any opinions, findings, and conclusions or recommendations expressed in this paper are those of the authors and do not necessarily reflect the views of the DOE.

\bibliographystyle{IEEEtran}
\bibliography{reference}

\end{document}